**SOFTWARE METAPAPER**

# Virtual Micromagnetics: A Framework for Accessible and Reproducible Micromagnetic Simulation

Mark Vousden, Marc-Antonio Bisotti, Maximilian Albert and Hans Fangohr
Faculty of Engineering and the Environment, University of Southampton, Southampton, SO16 7QF, GB
Corresponding author: Mark Vousden (mark.vousden@soton.ac.uk)

Computational micromagnetics requires numerical solution of partial differential equations to resolve complex interactions in magnetic nanomaterials. The Virtual Micromagnetics project described here provides virtual machine simulation environments to run open-source micromagnetic simulation packages [1]. These environments allow easy access to simulation packages that are often difficult to compile and install, and enable simulations and their data to be shared and stored in a single virtual hard disk file, which encourages reproducible research. Virtual Micromagnetics can be extended to automate the installation of micromagnetic simulation packages on non-virtual machines, and to support closed-source and new open-source simulation packages, including packages from disciplines other than micromagnetics, encouraging reuse. Virtual Micromagnetics is stored in a public GitHub repository under a three-clause Berkeley Software Distribution (BSD) license.

**Funding Statement:** We acknowledge financial support from EPSRC's DTC grant EP/G03690X/1.

## (1) Overview
### 1.1 Introduction
Micromagnetics is the study of the physical properties and behavior of magnetic materials at micrometer length scales. Micromagnetics research has promising applications in data storage and processing device design [2, 3]. The dynamics of micromagnetic systems are governed by competing energy terms, resulting in a complex system that can only be resolved using numerical simulation [4]. Several open-source numerical simulation software packages exist to simulate micromagnetic systems, each suitable to solve different problem types. Some of these packages are actively used and heavily cited in the literature [5] demonstrating their importance to micromagnetic research.

Since micromagnetic simulation packages are typically designed to run on a specific operating system, the user's choice of operating system may limit the number of simulation packages available to them. Even on a supported operating system, it is often difficult to install these simulation packages due to their complicated install procedures. This problem is worsened where the target audience of these simulation packages are researchers with focus in design, experimental, or analytical work who may not have computational science training. These factors inhibit the *accessibility* of the simulation software.

Another difficulty relates to the *reproducibility* of simulation results: if a result is computed with a particular simulation software, it should be reproducible in the future if the same simulation software is used again. However, when using the same version of the simulation software, different results may be computed if a different version of a third-party library is used [6].

Both problems of accessibility and reproducibility can be overcome by supplying pre-packaged execution environments in which micromagnetic simulations can be run, including third-party libraries and an operating system (providing accessibility). These environments remove the need for the user to deal with complicated installation procedures, preserve the simulation setup, and allow the researcher to share the environment with other users (providing reproducibility).

The Virtual Micromagnetics project provides such virtual machines in which numerical micromagnetic simulations can be run. Virtual machines, as described in Section 1.2.1, are processes that emulate a machine with an operating system. These virtual machines operate in isolation from the machine hosting them. This isolation allows virtual machines to execute simulations on personal computers, high-throughput compute clusters, and cloud compute services identically. Since a virtual machine emulates a real machine, no additional knowledge of interfaces or micromagnetic simulation is required beyond the initial brief setup procedure of the virtual machine. Virtual Micromagnetics supports four open-source packages: OOMMF [7], Nmag [8], Magpar [9], and Fidimag [10], and has also been used with closed-source software to generate results for two research publications [11, 12].

### 1.1.1 Use cases

Virtual Micromagnetics contains two usable components: the virtual environments produced from the build process as described in Section 1.2, and the source files that install micromagnetic simulation packages in an automated manner. The target audience of these components are "users" and "power users" respectively, where:

- **Users** include researchers that wish to access micromagnetic simulation software and run reproducible simulations, as well as students learning micromagnetics that require a stable micromagnetic simulation environment.
- **Power users** include system administrators looking to install micromagnetic packages, and micromagnetic simulation software developers looking to make their software more accessible.

User use cases ($U_i$), which informed the design decisions of Virtual Micromagnetics as described in Section 1.2, include:

- $U_1$ To run a micromagnetic simulation on a computer with an operating system not supported by the simulation package.
- $U_2$ To distribute micromagnetic simulation and post-processing scripts and data to others for reproducibility and collaboration.
- $U_3$ To archive micromagnetic simulation software as a supplement for reports and publications to enable reproducibility.
- $U_4$ To run micromagnetic simulations on cloud computing resources or high throughput computing hardware, where installing the micromagnetic software is more difficult due to lack of administrator rights.

Additionally, power user use cases ($P_j$) include:

- $P_1$ To provide an environment for users to run micromagnetic simulations using either existing software or newly developed micromagnetic software from the power user.
- $P_2$ To validate a new micromagnetic simulation package by direct comparison with existing simulation packages. For example, the micromagnetic standard problems [13, 14, 15, 16] can be executed inside the virtual machine. The results computed by each package can be compared to conclude whether or not a new package achieves results consistent with other packages.
- $P_3$ To install micromagnetic simulation software to run natively on specialist hardware, such as high performance computing clusters.

### 1.2 Implementation and architecture

In this section, we explain our choice of virtualization tool with reference to the use cases presented in Section 1.1.1 by discussing virtual machines, Linux containers, and the implications of virtual machines for high-performance computing with Graphics Processing Units (GPUs) and coprocessors. We then define what Virtual Micromagnetics environments are and how they are built. Lastly, we describe the process of provisioning virtual machines with micromagnetic simulation packages and other software required by Virtual Micromagnetics users.

### 1.2.1 Virtual Machines and Simulation Distribution

In this work, we define a virtual machine as a software that imitates a machine with an operating system (OS). This is a "system virtual machine" as defined by the literature [17]. Virtual machines effectively decouple the simulation software running inside the them from the underlying host operating system. Since virtual machines can be operated on a variety of different hardware and software configurations, this enables software tasks to be undertaken within precisely defined environments on many different host configurations. Running a virtual machine is a process defined by a set of files (of the order of gigabytes in size in Virtual Micromagnetics) representing both a hard disk, and a description of the hardware that the machine emulates. The isolated nature of virtual machines make them ideal for research, as the simulation environment can be archived and reproduced, satisfying use case $U_3$ defined in Section 1.1.1. Virtual Micromagnetics virtual machines use an Ubuntu GNU/Linux OS, which users will need to interact with in order to run simulations.

The file-based nature of virtual machines makes it possible to duplicate and distribute them, satisfying use case $U_2$. Virtual machines can contain the scripts and environment used to run a simulation, and can be included in support of a publication in addition to simulation data. If the machine supporting a publication is made freely available, any researcher can download it and run the simulation. This ease of distribution also eases the introduction of new researchers to micromagnetic simulation using Virtual Micromagnetics, as they no longer require extensive software knowledge to install micromagnetics simulation packages on their machines. However, since the files that define virtual machines are large, users must consider requirements on their network when downloading these virtual machines. An alternative mechanism to distribute simulation software is to host the virtual machine on a cloud computing service [18], so that other researchers can reproduce the published simulation without the networking and compute overheads for a fee. These two distribution mechanisms can be provided simultaneously, giving the target researcher the choice of how they wish to reproduce the simulation.

### 1.2.2 Linux Containers

Linux containers [19] are another virtualization technology that supports execution of applications in isolated environments. Containers do not run their own OS, and instead use functionality from the host OS as a substitute. This design reduces the hard disk requirement of a container compared to the corresponding virtual machine, since a virtual machine must contain an OS. Furthermore, applications run in a container do so at near-native processing, memory, and disk performance [20]. This is unlike a virtual machine, which demands additional computational overhead from the host to run its OS. While it is possible to use virtual machines in traditional high-performance computing clusters running GNU/Linux, containers are a superior option in this case due to their lightweight nature. However, unlike containers, virtual machines can run in isolation from the host OS since virtual machines contain an OS of their own. This allows a virtual machine running a GNU/Linux OS to be used on a host machine running a different OS, such as a Windows OS. This satisfies the operating system requirements in use case $U_1$. This feature also satisfies the cloud computing use case $U_4$, where the OS can vary, making virtual machines a superior choice over containers in this case. Virtual machines have been used in this project as opposed to containers, but container support is planned for a later release of Virtual Micromagnetics to support more use cases.

### 1.2.3 Effect of Virtual Machine use on Simulation Performance

Tasks run in virtual machines exhibit worse CPU, I/O, and network performance than tasks run natively [21, 22], though virtual machine performance has improved considerably with virtualization technology [20]. Furthermore, the magnitude of this performance loss varies greatly with the software and hardware of the host machine. This reduced performance impacts simulation time, which demonstrates a compromise between the reproducibility and accessibility of a simulation with simulation execution time.

Coprocessors [23] and Graphics Processing Units (GPUs) are seeing increased use in high-performance computing applications [24, 25]. When compared to the specification of CPUs, GPUs contain considerably more processors, but these processors run at a lower clock speed and have less internal memory. This architecture makes them suitable for problems that parallelize well and contain many compute operations, but have low memory requirements. Simulating these problems with GPUs can considerably reduce simulation execution time, but this compromises the portability, reproducibility and accessibility of the simulation because a specific class of GPU is required to run optimally. Micromagnetics suits this problem description, so micromagnetic simulation packages have been produced for GPU architectures using finite-element [26], and finite-difference approaches [27, 28], including the open-source MuMax [29] which is commonly used for numerical simulation in the literature.

Two methods exist to implement GPU computing in virtual machines. Firstly, while it is possible to virtualize a GPU using a physical GPU at a performance cost [30], this cost is often not suitable for high-performance computing simulations. An alternative used in cloud compute services is to connect the virtual machine to a physical GPU cluster. This approach maintains the performance gain from using GPUs in simulation, but at the cost of portability, as the simulation can only be run on the cloud service and other machines with that GPU. Virtual Micromagnetics environments do not currently support GPU-enabled simulations.

### 1.2.4 Build process

The build process for creating a virtual machine image is presented in **Figure 1**. Virtual machine images are binary files that precisely define a virtual machine. These files include the contents of the hard disk of the virtual machine, as well as a specification for the machine. An image allows the user to create Virtual Micromagnetics machines as shown by the run process in **Figure 1**.

The build process is triggered by commanding "`make full`" in the root directory of the Virtual Micromagnetics source code repository. Comprehensive instructions to install software and execute the build process are in the documentation.[1] To begin, a virtual machine image (A in **Figure 1**) of a minimal Ubuntu GNU/Linux installation is downloaded using Vagrant [31]. Vagrant is an automation tool that interfaces with other software to deploy computational environments, such as virtual machines. VirtualBox is one of multiple software packages that Vagrant interfaces with, which can create, run, and destroy virtual machines [32]. The second step in the build process uses VirtualBox to create a virtual machine (B) from the image downloaded in step (A). Once the machine is created, Ansible [33] is used to install the simulation software on the virtual machine, test the software, and configure other file system requirements. This process is called provisioning, and is described further in Section 1.2.5. The final stage of the build process is to package the provisioned machine (C) into an image (D). This new image can be downloaded by others to create a copy of the provisioned virtual machine on their hardware. Virtual Micromagnetics environments are online at http://atlas.hashicorp.com/virtualmicromagnetics. Users access the virtual machine by commanding "`vagrant init virtualmicromagnetics/full; vagrant up`",[2] which downloads the image and creates and starts the user's virtual machine (E).

### 1.2.5 Provisioning using Ansible

In this work, provisioning a virtual machine is the action of installing the simulation software, testing it, and performing other file system configuration on the virtual machine. Tools that help with provisioning are either instruction-driven or state-driven. Instruction-driven provisioners are defined by a set of instructions, whereas state-driven provisioners are defined by a desired end state. A state-driven provisioner is more suitable for provisioning virtual machines, since the final state of the machine is desired, as opposed to the steps needed to create the virtual machine. The Bourne-Again Shell (Bash) is an example of an instruction-driven provisioner. In Virtual Micromagnetics, we use the state-driven provisioner Ansible due to its ease of testing, and superior modularity and error control. This is discussed further in Section 1.3. The advantages of Ansible come at the cost of using software with less widespread familiarity than Bash.

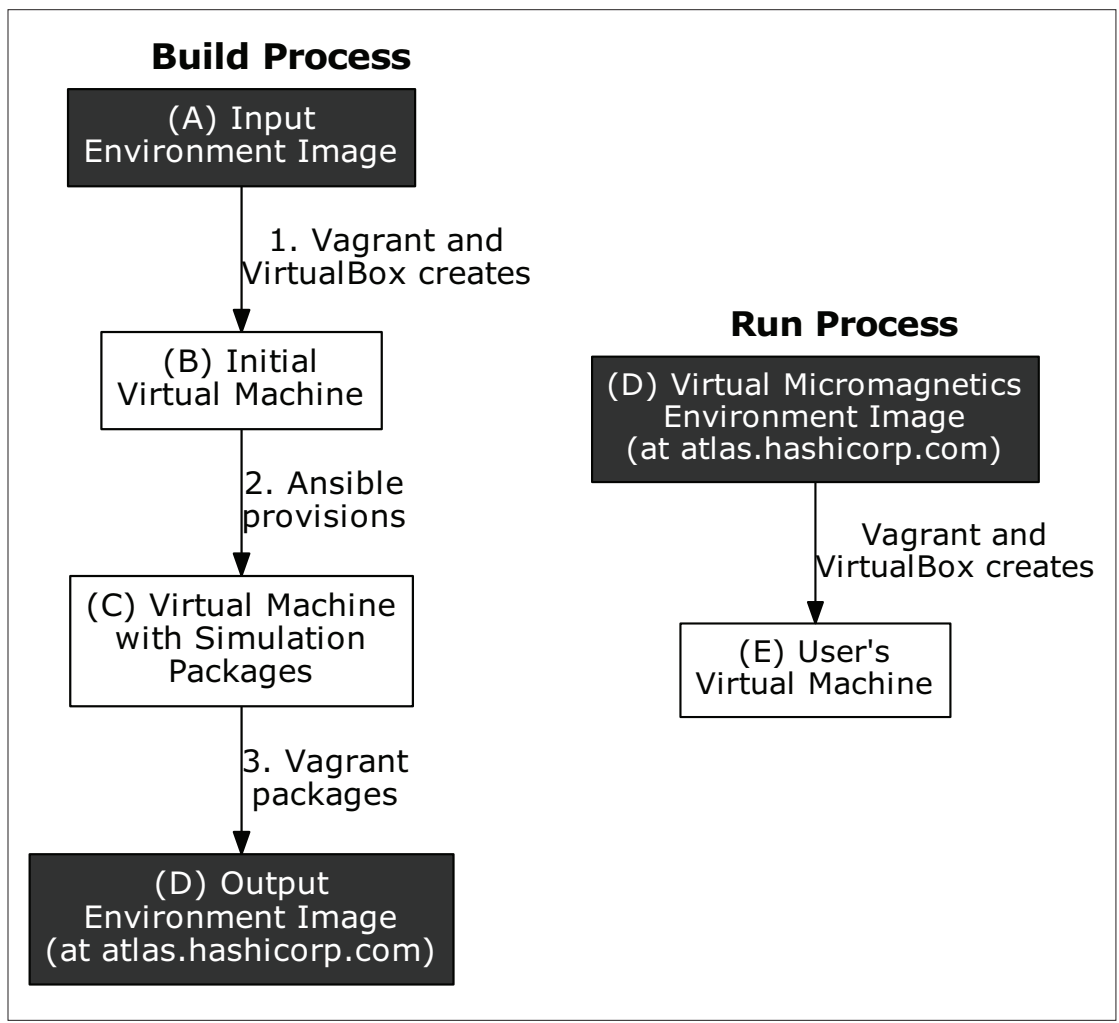

**Figure 1: Both processes:** Dark boxes represent virtual machine image files, whereas white boxes represent running virtual machines. **Left:** Automated build process flow. 1. Vagrant uses a virtual machine image (A) to create an initial virtual machine (B) using VirtualBox. 2. Ansible then provisions the virtual machine for simulation by installing simulation and support packages, and by altering the file system. 3. Vagrant then packages the virtual machine (C) as an image for users (D). **Right:** Automated run process flow. Vagrant downloads the image uploaded from the build process (D), and creates a virtual machine (E) from it for simulation, using VirtualBox.

Ansible allows easy customization of virtual machines by allowing the power user to install combinations of simulation software. For Virtual Micromagnetics, this process is described in the documentation[3], and means virtual machines can be built for individual needs; satisfying use case $P_1$ defined in Section 1.1.1. Furthermore, additional software installation procedures can be incorporated into this modular provisioning framework, meaning new private or public simulation software can be installed using Virtual Micromagnetics if the maintainer of that software provides Ansible scripts for the installation. The new software can be compared with existing software installed on the virtual machine, satisfying use case $P_2$. In addition, Ansible can also provision physical machines (as opposed to virtual machines), so Virtual Micromagnetics installation procedures can be applied to a variety of non-virtual systems, satisfying use case $P_3$. This feature allows power users to directly provision a machine in the cloud using Ansible.

### 1.3 Quality control

The virtual machines provided by Virtual Micromagnetics must be tested to ensure quality before they are distributed for research. To facilitate the automated execution of tests, a Jenkins [34] continuous integration server is used to build Virtual Micromagnetics virtual machine images.

Ansible, the software that provisions the virtual machine with simulation software, fails immediately when a step in the provisioning operation reports an error. This ensures that the virtual machine image produced by the build process (D) will only be produced if Ansible encounters no errors when provisioning a machine. This catches procedural errors, such as certain online resources being unavailable, as well as errors resulting from modifications made to the Virtual Micromagnetics source by developers or power users.

In addition to procedural errors, simulation package tests are also performed during the provisioning operation in the build process. These tests are written by the package maintainers to assure users that their software works once installed, and are used here to ensure simulation packages installed with Virtual Micromagnetics function correctly. Due to these tests, a machine with either faulty dependencies or a faulty installation of the simulation package will not be built. Micromagnetic simulation software tests are made accessible in Virtual Micromagnetics machines, so users can test each simulation package themselves for assurance. Details of simulation package tests can be found in the documentation of the simulation package that they test.

Prior to release, Virtual Micromagnetics virtual machine images produced by the build process are also checked by the following criteria:

- It creates a virtual machine without error when used with "`vagrant up`".
- Version and build information can be found on the machine, presently in the file "`virtualmicromagnetics_machine_characteristics.txt`" in the root directory, and on the desktop.
- Micromagnetic simulation software tests run successfully when run as a user on the virtual machine.
- Appropriate documentation is present on the Desktop of the virtual machine. These checks pre-empt software failures that would arise from the run process.

## (2) Availability
**Operating system**
To run Virtual Micromagnetics, any operating system supported by Vagrant 1.7.4 or greater and VirtualBox 5.0 or greater is supported by Virtual Micromagnetics. As of August 2016, this includes:[4]

- Windows Vista SP1+, 7, 8, 8.1, and 10 RTM build 10240 (32 and 64-bit).
- Windows Server 2008/R2, and 2012/R2 (64-bit).
- Mac OS X 10.8 – 10.11 (Intel hardware required) (64-bit).
- GNU/Linux Ubuntu 10.04 – 16.04, Debian 6.0, 8.0, and others (32 and 64-bit).

To build the Virtual Micromagnetics environments from source, any GNU/Linux or Mac OS X operating system listed above that supports Ansible 1.9 is supported.

**Programming language**
No installed programming language is needed to run Virtual Micromagnetics environments. To build these environments from the Virtual Micromagnetics source, interpreters for the following languages are required:

- Python 2 (2.7 and higher)
- GNU Make

**Additional system requirements**
Creating and running Virtual Micromagnetics environments requires resources to support the guest operating system and the host operating system simultaneously. As such, the following requirements are in addition to those of the host operating system.

- 1 GHz processor
- 2 GB system memory
- 10 GB disk space

The machine specifications will define the scale of simulations that can be run in Virtual Micromagnetics environments.

**Dependencies**
The additional software requirements to build or run Virtual Micromagnetics environments are:

- Run (Users)
  - VirtualBox (5.0 or greater)
  - Vagrant (1.7.4 or greater)
- Build (Power users)
  - As above
  - Ansible 1 (1.9 or greater)

**List of contributors**
- Mark Vousden: Software developer, documentation and website creator.
- Maximilian Albert: Contributed to Ansible roles for open-source packages.
- Marc-Antonio Bisotti: Contributed to initial project Ansible infrastructure.
- Hans Fangohr: Project supervisor.

**Software location**
*Archive*
  **Name:** Zenodo
  **Persistent identifier:** https://doi.org/10.5281/zenodo.53870
  **Licence:** Three clause Berkeley Software Distribution (BSD)
  **Publisher:** Mark Vousden
  **Version published:** 1.0.3
  **Date published:** 03/05/2016

**Code repository**
  **Name:** GitHub
  **Persistent identifier:** https://github.com/computationalmodelling/virtualmicromagnetics
  **Licence:** Three clause Berkeley Software Distribution (BSD)
  **Date published:** 03/05/2016 (Version 1.0.3)

**Other resources**
  **Homepage:** http://virtualmicromagnetics.org
  **Documentation:** http://virtual-micromagnetics.readthedocs.io

**Language**
YAML [35] is the primary language used in this repository. It is used by Ansible to both create and define the state of software in Virtual Micromagnetics environments.

## (3) Reuse potential
Section 1.1 highlights use cases for the Virtual Micromagnetics project. The number of potential users of this technology is driven by the number of researchers working with micromagnetic simulation. The list of publications using and citing the OOMMF micromagnetic simulation package alone shows a volume of over two-thousand published papers [5]. This provides a lower limit of scientific and engineering work that can benefit from the Virtual Micromagnetics project.

While Virtual Micromagnetics was developed within the computational micromagnetics community, its modular design makes it applicable to software outside of micromagnetic simulation. Virtual Micromagnetics can be used to create virtual environments containing different software for other communities, simply by specifying different Ansible roles as described in Section 1.2.5. This means Virtual Micromagnetics can be used for similar projects in other disciplines, as it is not limited to micromagnetic research in any way. Reproducibility and accessibility is a goal in all disciplines running numerical simulation, hence this wide audience should assure widespread reuse of Virtual Micromagnetics.

Interdisciplinary reuse of Virtual Micromagnetics is encouraged by the developers. Support for Virtual Micromagnetics exists for users through the GitHub repository issue tracker.

**Competing Interests**

The authors declare that they have no competing interests.

**Notes**

1. http://virtual-micromagnetics.readthedocs.io/en/1.0.3/getting-started-poweruser.html
2. http://virtual-micromagnetics.readthedocs.io/en/1.0.3/getting-started-user.html
3. http://virtual-micromagnetics.readthedocs.io/en/1.0.3/developer-notes.html
4. https://www.virtualbox.org/manual/ch01.html#hostossupport